\documentclass[reprint,superscriptaddress,amsmath,aps,showpacs,pra]{revtex4-2}
\usepackage{amssymb}
\usepackage{appendix}
\usepackage{verbatim}
\usepackage{float}
\usepackage{color}
\usepackage{textcomp}
\usepackage{gensymb}
\usepackage{dsfont}
\usepackage{hyperref}
\hypersetup{
     colorlinks   = true,
     citecolor    = blue
}
\usepackage{siunitx}
\usepackage{graphicx}
\usepackage{dcolumn}
\usepackage{bm}
\usepackage{braket}
\usepackage{empheq}
\usepackage{nicefrac} 
\usepackage[dvipsnames]{xcolor}
\renewcommand{\vec}[1]{\bm{#1}}

\RequirePackage{color}
\usepackage[normalem]{ulem}

\begin{document}

\title{Designing Band Structures by Patterned Dielectric Superlattices}
\author{Zhen Zhan}
\email{zhenzhanh@gmail.com}
\affiliation{Imdea Nanoscience, Faraday 9, 28015 Madrid, Spain}
\author{Yonggang Li}
\affiliation{Key Laboratory of Artificial Micro- and Nano-structures of Ministry of Education and School 
of Physics and Technology, Wuhan University, Wuhan 430072, China}
\author{Pierre A. Pantale\'on}
\affiliation{Imdea Nanoscience, Faraday 9, 28015 Madrid, Spain}
\date{\today}

\begin{abstract}

We investigate the electronic structure of graphene monolayers subjected to patterned dielectric superlattices.\ Through a quantum capacitance model approach, we simulate realistic devices capable of imposing periodic potentials on graphene.\ By means of both tight-binding and continuum models, we analyze the electronic structure across varied patterning geometries, including triangular, kagome, and square configurations.\ We explicitly explore the influence of device parameters such as the superlattice potential strength, geometry, and periodicity on the electronic properties of graphene.\ By introducing a long-range Coulomb interaction, we found an emergent periodic potential strong enough to open a mass gap, thereby generating a Chern band.\ Our study highlights the robustness and versatility of patterned dielectric superlattices for band engineering in graphene systems.

\end{abstract}

\maketitle
\section{Introduction}

The discovery of unconventional superconductivity in twisted~\cite{cao2018unconventional,park2021tunable,park2022robust} and non twisted~\cite{Zhou2022isospin,Zhang2022enhanced,Holleis2023Ising,Zhou2021superconductivity} graphene stacks has attracted a lot of interest. The possibility of modifying their electronic properties with the carrier density and external fields has started a new era in materials science~\cite{andrei2020graphene}.\ Experimentally, the graphene stacks are usually supported on a substrate which is typically a thin sample of hexagonal boron nitride (hBN)~\cite{Wang2019new}.\ Interestingly, numerous reports have suggested that the hBN substrate plays a pivotal role in various phenomena observed in graphene moir\'e systems, including Chern and Mott insulating states~\cite{wang2015evidence}, orbital magnetization~\cite{Sharpe2021Evidence}, valley Hall effects~\cite{Aktor2021Valley,Gorbachev2014Detecting}, ferroelectricity~\cite{zheng2020unconventional}, non-linear Hall effects~\cite{he2022graphene} and other correlated states~\cite{chen2019signatures}.\ It is well-established that placing graphene layers on a hBN substrate creates a periodic moir\'e-like superlattice (SL) potential~\cite{Yankowitz2014Graphene,Moon2014Electronic}.\ This SL potential, acting on the graphene carriers, results from both the mismatch between the lattice constants of hBN and the graphene layers, and the different rotation between these two layers~\cite{Jung2015Origin}.\ The strength and periodicity of the moir\'e-like SL potential can be simultaneously modified by the degree of alignment or by encapsulation, while the symmetry is fixed by the substrate~\cite{dean2010boron,Hunt2013Massive,Song2013Electron,Amet2013Insulating,Chen2014Observation,Wong2015Local,Jung2015Origin,Lee2016Ballistic,Wang2016Gaps,Yankowitz2018Dynamic,Zibrov2018Even,Kim2018Accurate,Jung2015Origin}. A moir\'e-like SL potential can also be generated by moir\'e localized states. For instance, the optically excited Rydberg excitions in monolayer tungsten diselenide could be confined and controlled via a remote Coulomb SL generated by the localized states in a small-angle twisted bilayer graphene \cite{Hu2023}.\ Moreover, correlated insulators in bilayer graphene are engineered via a remote Coulomb SL realized by localized states in twisted bilayer WS$_2$~\cite{Zhang2024}.

An alternative way to generate a SL potential is through an electrostatic gating scheme, where an artificial structure is designed by patterning an atomically thin van der Waals material~\cite{forsythe2018band}.\ Unlike moiré-like SL potentials induced by thin substrates, artificial SL potentials offer the advantage of flexible control over lattice patterns, symmetry, and \textit{in-situ} control of the potential strength~\cite{forsythe2018band,huber2020gate,li2021anisotropic}. On the experimental side, significant progress has been made in creating patterned structures with various geometries, such as one-dimensional arrays~\cite{li2021anisotropic,drienovsky2018commensurability}, square~\cite{forsythe2018band,huber2020gate,barcons2022engineering}, triangular~\cite{barcons2022engineering}, and recently, kagome lattices~\cite{Wang2024Dispersion}. Depending on the experimental technique, the patterned gates are designed with lattice periods ranging from 80~\cite{Wang2024Dispersion} to 20 nm~\cite{barcons2022engineering}. Specially, the most common patterning technique is electron beam lithography (EBL), which can create patterns with a minimum size of 35 nm pitch~\cite{jessen2019lithographic}. Another common approach is the focused-ion beam (FIB) milling, which can pattern 2D SLs with periods down to sub-20 nm~\cite{barcons2022engineering}.

Previous analytical work predicted that when a periodic potential is applied to a graphene monolayer, additional massless Dirac fermions are formed at the edges of the induced superlattice Brillouin zone (sBZ)~\cite{park2008new,barbier2010extra,Guinea2010Band}. Further studies, owing to advancements in nanopatterning techniques, observed the replica of the Dirac cones, Landau fans, Hofstadter spectrum, and Brown-Zak oscillations~\cite{forsythe2018band,huber2020gate,li2021anisotropic,barcons2022engineering,huber2022band}, effectively reproducing behaviors seen in graphene/hBN moiré systems~\cite{ponomarenko2013cloning}. The electronic structures and transport properties of graphene systems with a SL potential have been studied using the continuum model~\cite{ghorashi2023topological,krix2023patterned,ghorashi2023multilayer,zeng2024gate,hougland2021theory} and a combination of a scalable tight-binding (TB) model with the real-space Green's function approach~\cite{chen2020electrostatic,mrenca2022quantum,mrenca2023probing}. Aside from monolayer graphene, artificial SL potentials can also be applied to other systems, including transition metal dichalcogenides (TMDCs)~\cite{shi2019gate,yang2022chiral,garcia2024fractal}, multilayer graphene~\cite{ghorashi2023topological,sun2023signature,krix2023patterned,ghorashi2023multilayer,zeng2024gate}, and gallium arsenide (GaAs) quantum wells~\cite{singha2011two,wang2024lateral,wang2024tuning}. However, a comprehensive analysis on its feasibility, considering experimentally accessible parameters, and the role of electron-electron interaction is still lacking.

\begin{figure*}
    \centering
    \includegraphics[width=0.8 \textwidth]{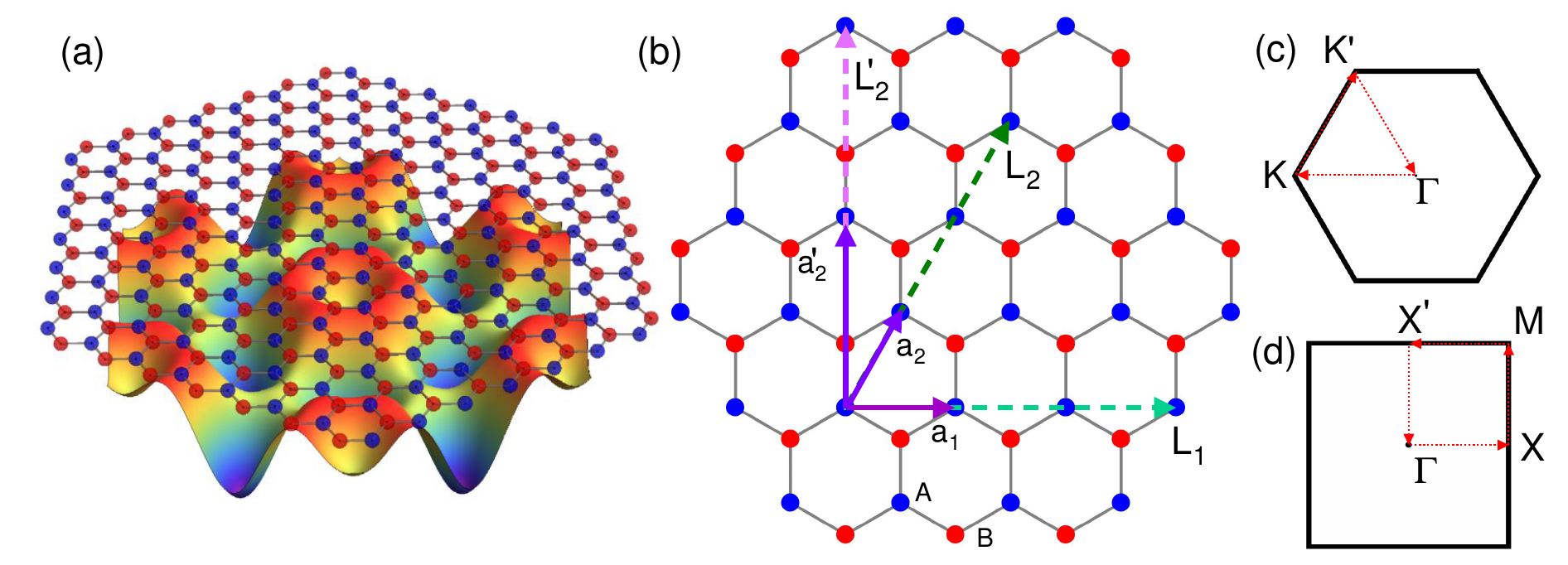}
    \caption{(a) Illustrative representation of a graphene monolayer with a periodic potential. The A and B sites of the honeycomb lattice are indicated with different colors. 
    (b) Real space lattice structure. The periodic potential unit cell is constructed using the primitive lattice vectors $\vec{a}_1$ and $\vec{a}_2$ of graphene. For a triangular SL we have $\vec{L}_1=n a_1$ and $\vec{L}_2=n \vec{a}_2$ (light and dark green dashed lines, respectively). For a rectangular SL we have $\vec{L}_1=n \vec{a}_1$ and $\vec{L}'_2=m \vec{a}'_2$ with $\vec{a}'_2 = 2\vec{a}_2-\vec{a}_1$ and $n,m$ integers. In (c) we display the hexagonal sBZ for both kagome and triangular SL. In (d) we display the square sBZ. The path with arrows is used in the numerical calculations. For visualization purposes, in (b) we set $n=3$ for a triangular and $n=3$, $m=2$ for a rectangular SL. Note that, for large values of $m$ and $n$, we can approximately define a square SL. 
    }
    \label{fig:FigLattice}
\end{figure*}

In this paper, we explicitly investigate the electronic structure of a graphene monolayer with a patterned dielectric lattice (Fig. \ref{fig:FigLattice}(a)). Using a quantum capacitance model approach, we simulate realistic patterned devices that generate a periodic potential acting on a graphene monolayer. We employ both TB and continuum models to explore the electronic structure with patterning in three different geometries: triangular, kagome, and square lattices. We explicitly investigate the effects of various tuning knobs of the device, including the strength of the SL potential, lattice geometry, and period, on the electronic structures. We also study the evolution of the density of states (DOS) with the lattice strength and describe an emergent periodic potential from purely electron-electron interactions. Our work demonstrates that the patterned dielectric superlattice is a robust and versatile technique for band engineering of graphene systems. The TB and continuum models agree well within certain ranges of parameters, providing efficient tools to investigate open questions regarding these new systems.


This paper is organized as follows. In Sec. \ref{sec: Theoretical}, we derive the lattice structure and the TB model for graphene with a SL potential. The calculation of the realistic SL potential using the Finite Element Method (FEM) is detailed in Sec. \ref{potential sec}. In Sec. \ref{continuum}, we descript the effective continuum model for the graphene with SL potentials. On the basis of the TB and continuum models, we study the electronic structures of graphene with SL potentials, in particular, the effects of the potential shape, strength, period and electron-electron interactions, in Sec. \ref{results_sec}. Finally, conclusions are given in Sec. \ref{conclu}. 

\section{The tight-binding Method} \label{sec: Theoretical}
\subsection{Lattice Structure} \label{subsec:Lattice}

We consider a monolayer graphene, whose primitive lattice vectors are $\vec{a}_{1}=a\left(1,0\right)$ and $\vec{a}_{2}=a\left(1/2,\sqrt{3}/2\right)$, where $a$ is the lattice constant ($a\simeq2.46\:\textrm{Å}$ in graphene).\ The reciprocal lattice vectors satisfy $\vec{a}_{i} \cdot \vec{b}_{j} = (2\pi)\delta_{ij}$ with $\vec{b}_1=(4\pi/\sqrt{3}a) \left(\sqrt{3}/2,-1/2\right)$ and $\vec{b}_2=(4\pi/\sqrt{3}a) \left(0,1\right)$.\ As shown in Fig.~\ref{fig:FigLattice}(b), the graphene honeycomb lattice is bipartite with $A$ and $B$ sites, and their coordinates can be obtained with 
\begin{equation}
    \vec{r} = i \vec{a}_1 + j \vec{a}_2 + s \vec{\delta}_{ab},
    \label{eq: LatticeSites}
\end{equation}
where $\vec{\delta}_{ab} = (\vec{a}_1+\vec{a}_2)/3$ is a shift between A and B sites, with ${i,j}$ integers and $s\in\{0,1\}$ the sublattice index. We define a SL potential $v_{SL}$ as a smooth periodic function acting on each of the graphene lattice sites.\ That is, $v_{SL} (\vec{r}+\vec{L}_1)=v_{SL} (\vec{r}+\vec{L}_2)=v_{SL} (\vec{r})$, with $\vec{L}_1$ and $\vec{L}_2$ as the primitive lattice vectors of the SL.\ In the case of a triangular SL, green dashed lines in Fig.~\ref{fig:FigLattice}(b), we have $\vec{L}_1=n \vec{a}_1$ and $\vec{L}_2=n \vec{a}_2$, where $n$ is an integer, and the SL length is given by $L_m = |\vec{L}_1|=|\vec{L}_2|= n |\vec{a}_{1,2}|$.\ The reciprocal lattice vectors are $\vec{G}_{1,2} = (1/n) \vec{b_{1,2}}$.\ 
The kagome SL can be constructed by using the triangular SL basis vectors, however, in this case there are three different periodicities due to the corner shared triangles, see Fig.~\ref{fig: Onsite}.\ For a square SL, we have $\vec{L}_1=n \vec{a}_1$ and $\vec{L}'_2=m \vec{a}'_2$, where $\vec{a}'_2 = 2\vec{a}_2-\vec{a}_1$ and $n$ and $m$ are integers, light green and purple dashed arrows in Fig.~\ref{fig:FigLattice}(b).\ The reciprocal lattice vectors are then given by $\vec{G}_{1} = (4\pi/\sqrt{3}a) \left(\sqrt{3}/2n,0\right)$ and $\vec{G}_{2} = (4\pi/\sqrt{3}a) \left(0,1/2m\right)$.\ 
Note that due to the honeycomb lattice, a perfect square SL cannot be strictly defined, and in general $|\vec{L}_1| \neq |\vec{L}'_2|$. However, for a large SL period with $L_m\gg a$, a nearly square SL can be defined by adjusting the $m, n$ indexes. 
The sBZ is shown in Fig.~\ref{fig:FigLattice}(c) for a triangular and kagome SLs. In Fig.~\ref{fig:FigLattice}(d) we display the sBZ for a square SL. Red dashed arrows in both panels indicate the path used in the numerical diagonalization. 


\subsection{Tight-binding Model} \label{sec: TB Model}

We describe an atomistic TB model used to calculate the electronic structures of graphene with a SL potential. We begin with a general expression of the TB model for multilayer graphene systems, considering only the $ p_z$ orbitals of carbon atoms, which is sufficiently accurate in the low-energy range. We write the Hamiltonian in the basis orbitals in the language of wave functions ($|i\rangle$) as, 
\begin{equation}
H=\sum_{i}\epsilon_i|i\rangle\langle i|+\sum_{\langle i,j \rangle} t_{ij} | i \rangle \langle j| +\sum_i v_{SL}^i |i\rangle\langle i|,
\label{eq: H_tb}
\end{equation}
where $\epsilon_i$ is the on-site potential of the $i$-th orbital and $\langle i,j \rangle$ is the sum on index $i$ and $j$ with $i\neq j$. $t_{ij}$ is the coupling matrix element between two $p_z$ orbitals located at $\mathbf{r}_i$ and $\mathbf{r}_j$, which is defined as, 
\begin{equation}
t_{ij}=n^2V_{pp\sigma}(r_{ij})+(1-n^2)V_{pp\pi}(r_{ij}),
\label{eq: tij}
\end{equation}
where $n=\frac{z_{ij}}{r_{ij}}$ is the direction cosine of $\vec{r}_{ij}=\vec{r}_j-\vec{r}_i$ along the direction perpendicular to the graphene plane and $r_{ij}=|\vec{r}_{ij}|$. The hopping parameters follow a distance-dependent Slater-Koster format given by
\begin{align}
  V_{pp\pi}(r_{ij})= -\gamma_0 e^{ 2.218 (a_0 - r_{ij})}F_c(r_{ij}),\\
  V_{pp\sigma}(r_{ij})=\gamma_1 e^{ 2.218 (d - r_{ij})}F_c(r_{ij}).
\end{align}
Here, $d$ is the interlayer distance, $a_0 = a/\sqrt{3}$ is the carbon-carbon bond length, and $\gamma_0$ is the first-neighbor intralayer interaction, which is related to the Fermi velocity $v_F \approx 3a_0\gamma_0/(2\hbar)$. The parameter $\gamma_1$ represents the interlayer interaction between two $p_z$ orbitals in a $\pi$ configuration. The value of $\gamma_1$ is obtained by fitting with a DFT calculation around the Dirac energy in AA and AB stackings. In the monolayer graphene case, we only consider the intralayer interactions, the $pp\pi$ terms in Eq.~(\ref{eq: tij}). In this paper, we choose the hopping parameter as $\gamma_0 = 2.7$~eV. The smooth function is $F_c(r) = [1 + e^{(r-r_c)/l_c}]^{-1}$, in which $r_c$ and $l_c$ are chosen as 6.14 and $0.265 \text{\AA}$, respectively.

\begin{figure*}[t]
    \centering    
    \includegraphics[width=0.9\textwidth]{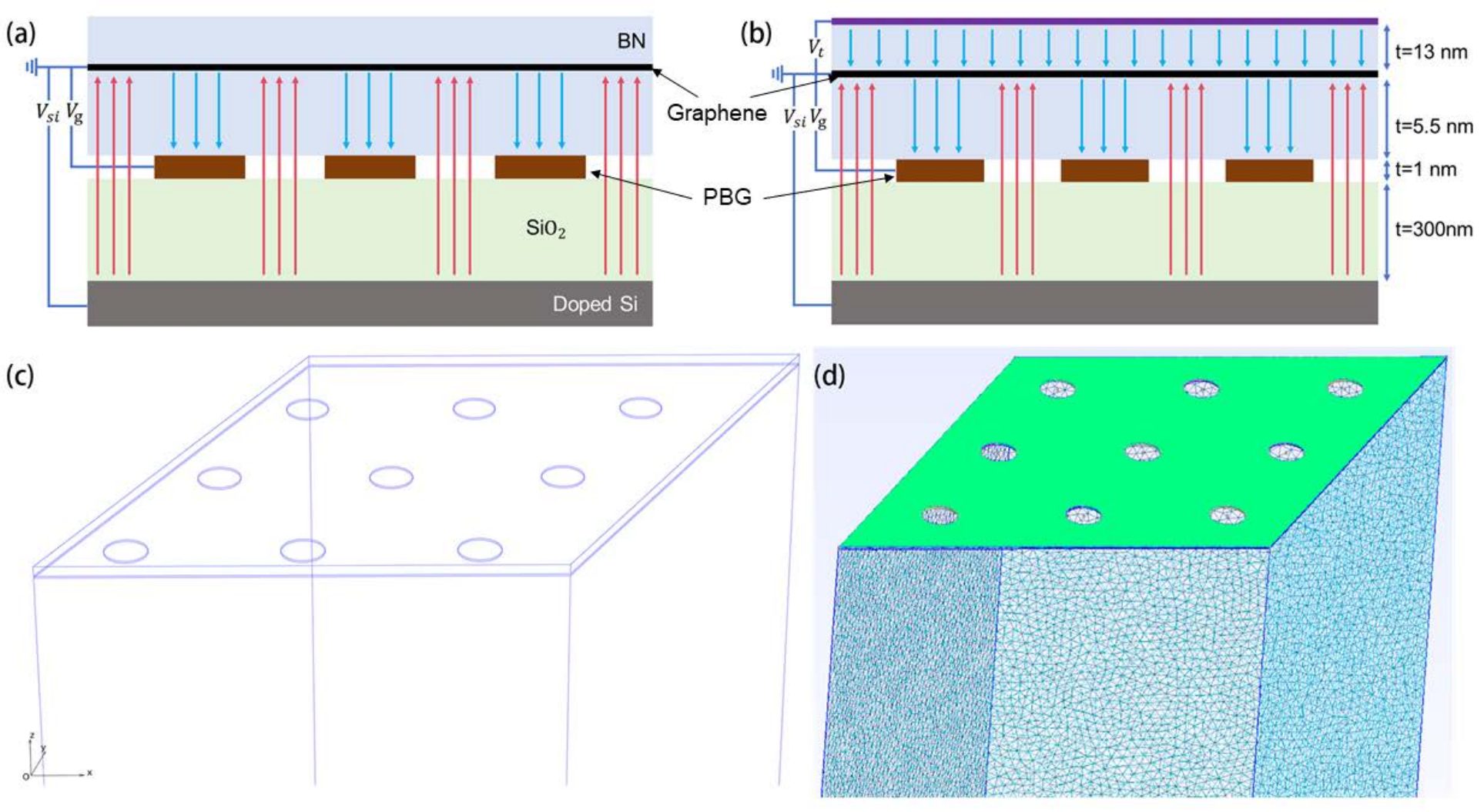}
    \caption{In (a) and (b) we illustrate two-gate and three-gate experimental setups which allow for a tunable and spatially varying  potential. In (c) we show the geometry of (a) for the electrostatic simulation. We use the triangular patterning as an example. In (d) we show an example of the mesh generated by the GMSH algorithm. This mesh is used to calculate the spatial variation of the electronic density.}
    \label{fig:Device}
\end{figure*}

In Eq.~(\ref{eq: H_tb}), the term $v_{SL}(\vec{r}_i)$ represents the SL potential introduced by the bottom electrostatic gate (see Fig.~\ref{fig:Device}).\ This site-dependent gate potential can be obtained from a finite-element-based electrostatic simulation, as we will describe in the following section.\ Our graphene TB Hamiltonian in Eq.~(\ref{eq: H_tb}) consists of an onsite term $\epsilon_i$ and a distance-dependent intralayer hopping term $t_{ij}$ with a cutoff of 0.5 nm.\ Note that if we only consider the nearest-neighbor intralayer hopping in Eq.~(\ref{eq: H_tb}), the intrinsic Dirac cone of the monolayer graphene will always have energy $E_D=0$, and the on-site potential $\epsilon_i$ is zero.\ In this paper, the intralayer coupling between atoms extends beyond the first nearest neighbors, which breaks the electron/hole symmetry of the monolayer graphene. Consequently, the energy $E_D$ is not zero if we reset the origin of the energy to the intrinsic Dirac point of monolayer graphene.\ Additionally, the SL potential $v_{SL}$ introduces a global shift of the band structure. Therefore, to reset $E_D=0$ in the band structure, different shift energies are required for different SL potentials.

In TB calculations with a patterned gate we need to define a unit cell to perform the numerical calculations. The size of the unit cell will depend on the symmetry and periodicity of the patterning~\cite{hougland2021theory}.\ In particular, we choose SL potentials such that they satisfy the condition $v_{SL} (\vec{r}+\vec{L}_1)=v_{SL} (\vec{r}+\vec{L}_2)=v_{SL} (\vec{r})$, with $\vec{L}_1$ and $\vec{L}_2$ linear combinations of the graphene primitive lattice vectors.\ In the cases considered in this work, the periodicity of both triangular and kagome potentials satisfy the previous condition.\ As illustrated in Fig.~\ref{fig:FigLattice}(b), the period $\vec{L}_1$ (or $\vec{L}_2$) of these potentials can be written as a multiple of the graphene primitive lattice vectors, this is $\vec{L}_{1,2} = n \vec{a}_{1,2}$. If $n$ is a multiple of three, the graphene intrinsic Dirac point maps onto the $\Gamma$ point of the sBZ. Otherwise, the graphene intrinsic Dirac point maps onto the $\vec{K}$ or $\vec{K'}$ corners.\ The square potential is different from the above hexagonal lattice potentials.\ In the graphene lattice, it is not possible to find four points to form a square. We select the primitive lattice vectors to be $\vec a_1$ and $\vec a_2'$, as shown in Fig.~\ref{fig:FigLattice}(b).\ The size of the unit cell is $a\times \sqrt{3}a$.\ The period of the square potential in each direction is chosen as $L_1=na_1$ and $L_2'=ma_2'$. In order to map the graphene Dirac point onto the high-symmetry point $\Gamma$ of the sBZ in square lattice, the integer $n$ is a multiple of three. Otherwise, the graphene intrinsic Dirac point will be mapped onto other points of the sBZ.   

The band structure of the graphene SL is calculated by directly diagonalizing the Hamiltonian in Eq.~(\ref{eq: H_tb}).\ Note that in the current state of the art, micro-scale areas with minimum size down to 8 nm could be patterned via the focused-ion beam milling approach, providing a patterned gate with a period as small as 16 nm~\cite{barcons2022engineering}.\ For a triangular SL with that size, the number of atoms in a unit cell is about nine thousand, of which the diagonalization process is extremely time-consuming.\ Therefore, we use a rescaling method to calculate the band structure~\cite{liu2015scalable} where low-energy electronic structure of the graphene SL can be reproduced with a smaller number of atoms.\ The scaled SL keeps the Fermi velocity and SL period invariant. We use the following transformations,
\begin{align}
\label{scale_TB}
a_0'\to s_f a_0, \\\nonumber
\gamma_0' \to \frac{1}{s_f}\gamma_0,\nonumber
\end{align}
to scale the TB Hamiltonian in Eq.~(\ref{eq: H_tb}). The validity criterion of the scaling approach is \cite{liu2015scalable}:
\begin{equation}
    s_f \ll \frac{3\gamma_0 \pi}{|E_{max}|}
\end{equation}
where $E_{max}$ is the maximal energy of interest in the calculations. With an appropriate $s_f$, we can calculate the band structure of graphene SL with period up to 80 nm. In this work, we set $s_f=5$. All the TB calculations are performed in the TBPLaS simulator \cite{li2023tbplas}.


\section{Patterned dielectric potentials}
\label{potential sec}

In this part, we describe a methodology to numerically obtain a realistic potential profile by considering the different geometries of the patterned dielectric gate. Figures~\ref{fig:Device}(a) and (b) show two typical experimental setups. For the two-gate device in Fig.~\ref{fig:Device}(a), an artificial SL potential is created under the synergistic control of the patterned bottom gate $V_g$ and the global back gate $V_{si}$. The gate $V_{si}$ manipulates the strength of the SL potential acting on the graphene layer.\ The gate $V_g$ is used to independently tune the carrier density in the graphene channel.\ On the other hand, in the three-gate device shown in Fig.~\ref{fig:Device}(b), a global top gate $V_t$ is applied on the top of the device, which in combination with the gates $V_g$ and $V_{si}$ provides a tunable artificial SL potential. Note that the top gate $V_t$ can uniformly tune the carrier density in the graphene channel, whereas the gate $V_g$ can only efficiently tune the carrier density in the anti-hole sites of the graphene layer~\cite{Wang2024Dispersion,barcons2022engineering}.\ When $V_t = 0$, the devices in Figs.~\ref{fig:Device}(a) and (b) are equivalent.\ Moreover, up to date, the gate-controlled SL potential could be obtained by two different methods. In the first method, between the hBN/graphene/hBN stack and the SiO$_2$ substrate, a periodic bottom gate (PBG) made by patterning a few-layer graphene is inserted (Figs. \ref{fig:Device}(a) and (b)).\ Due to the presence of the hole sites in PBG, we could achieve a gate-controllable SL potential.\ This method has been experimentally realized in Refs.~\cite{huber2020gate,Wang2024Dispersion}. In the second method, the hBN/graphene/hBN is directly placed on a spatially periodic prepatterned SiO$_2$ substrate with user-defined geometry. In this case, the SL potential is controlled by the global bottom gate $V_{si}$, and the carrier density in the graphene channel is uniformly tuned by the top gate $V_t$.\ This method has been experimentally demonstrated in Refs.~\cite{forsythe2018band,li2021anisotropic}.\ The SL potential profile introduced in graphene by the SL gate could be obtained by using a commercially available electrostatic modeling package COMSOL, or by the open-source packages FENICS~\cite{logg2012automated} and GMSH~\cite{geuzaine2008gmsh, Geuzaine2009} based on the FEM method. In the following, we will give an explicitly description how to estimate the realistic SL potential via the FEM method.

\begin{figure*}[ht]
\centering    
  \includegraphics[width=\textwidth]{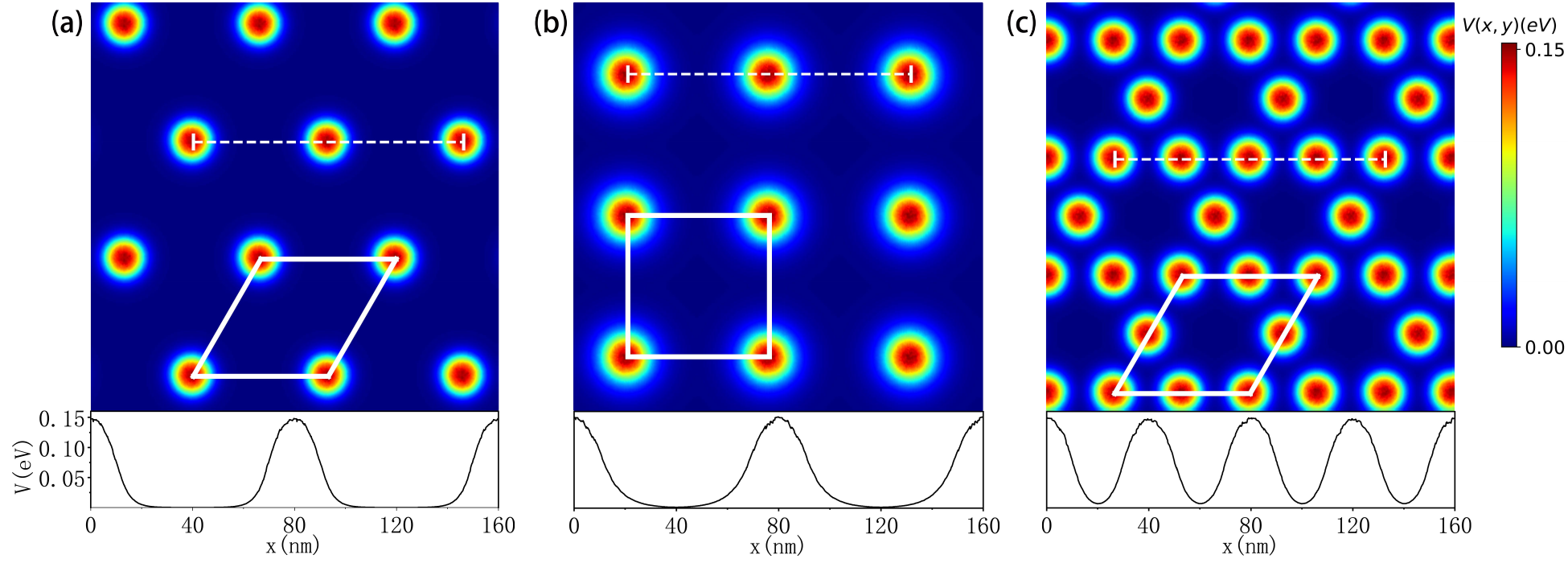}
  \caption{(a)-(c) The calculated potential profiles in the x-y plane (upper panel) and in the x direction (lower panel) of the triangular, square and kagome SL cases, respectively. The corresponding line profiles are outlined with white dashed line in the lower panel. The unit cell of the patterned SL is illustarted by the white solid quadrangle. The three SL potentials have the same length $L_1=80$ nm, and SL gate bias $V_{si}=-80$ V.}
\label{fig: Onsite}
\end{figure*}

We consider the device in Fig.~\ref{fig:Device}(a) as an example to estimate the SL potential acting in a single-layer graphene placed in the $z=0$ plane.\ To obtain realistic onsite energy profiles $v_{SL}(x,y)$, we need to know the carrier density $n(x,y)$, which could be solved within a quantum capacitance model~\cite{liu2013theory}.\ In this situation, the combined system graphene/substrate/gates can be considered as a parallel-plate capacitor, and the carrier density in graphene as the surface carrier density induced by the gates.\ We first build a three-dimensional (3D) electrostatic model and generate a finite-element mesh using GMSH~\cite{geuzaine2009gmsh,Geuzaine2009} shown in Figs.~\ref{fig:Device}(c) and (d), respectively.\ Then, we obtain the electrostatic potential $U(x,y,z)$ in the model by solving the Poisson equation using FENICS (automated partial-differential equation solver)~\cite{logg2012automated}.\ In solving the Poisson equation, non-zero constant boundary conditions $V_t$, $V_g$, and $V_{si}$ are set at the gates, as shown in Fig.~\ref{fig:Device}, while the graphene is grounded throughout, maintaining a potential of zero.\ Additionally, periodic boundary conditions are applied in the $x/y$ direction of the device.\ Next, the surface charge density in the graphene is obtained as
\begin{equation}
    \sigma(x,y) = \left. -\epsilon_r\epsilon_0\frac{\partial U(x,y,z)}{\partial z} \right|_{z=0},
    \label{eq:sigma}
\end{equation}
where $\epsilon_0$ is the permittivity in free space and $\epsilon_r$ is a position-dependent relative permittivity.\ The gate capacitance (considering $V_{si}$ as an example) is defined as 
\begin{equation}
    C_{si} = \left. \frac{\sigma(x,y)}{V_{si}} \right|_{V_t=V_g=0}.
    \label{eq:Capacitance}
\end{equation}
The above equation indicates that, when solving for the capacitance of a particular gate, the other gates need to be grounded.\ The classical carrier density is defined as,
\begin{equation}
n_c(x, y)=(C_{t}/e)V_{t}+(C_g/e)V_g+(C_{si}/e)V_{si}
\label{eq:CapacitanceDensity}
\end{equation}
with the gate capacitance calculated using Eq.~(\ref{eq:Capacitance}).\ Due to the finite capacity of graphene to reside electrons, a quantum correction terms $n_Q$ is included in the spatial carrier density $n=n_c+n_Q$. The quantum correction term is given by~\cite{liu2013theory}
\begin{equation}
n_Q=\frac{\pi}{2}\Big( \frac{\hbar v_F}{e^2} (C_t+C_g+C_{si})\Big)^2.
\end{equation}
This correction is small for the cases considered here, and is thus ignored~\cite{Wang2024Dispersion}. For a monolayer graphene, within the linear Dirac model~\cite{liu2013theory}, the superlattice potential can be calculated from the surface carrier density as,   
\begin{equation}
v_{SL}(x,y)=-sgn[n(x, y)]\hbar v_F \sqrt{\pi |n(x,y)|}, 
\label{eq:DensityPotential}
\end{equation}
where $v_F$ is the Fermi velocity of graphene and $n$ is the total density.\ Finally, the position-dependent periodic potential obtained from Eq.~(\ref{eq:DensityPotential}) is adjusted to the graphene lattice as a site-dependent on-site potential in Eq.~(\ref{eq: H_tb}). It has been proven that the exact solution for the space-resolved carrier density within the quantum capacitance model is equivalent to the self-consistent Poisson-Dirac iteration method \cite{liu2013theory}. In Fig.~\ref{fig: Onsite}, from left to right, we display the resulting potentials of a triangular, square and kagome patterned gates, respectively.\ In the bottom of each panel we show the corresponding potential cut, indicated by a dashed white line in each density plot. The periodicity in all cases is set to $L_m = 80$ nm. We note that the capacitance model described in this section with the TB model from the previous section fully determine the electronic properties of a graphene monolayer under the effects of a periodic patterned gate with an arbitrary geometry. 

\section{Effective continuum model}
\label{continuum}

\subsection{Low Energy Hamiltonian}

We now construct a low-energy continuum model. The effect of the periodic patterned gate is represented by a modulated SL potential, as  described in Sec.~\ref{subsec:Lattice}, and shown in Fig.~\ref{fig: Onsite} for different patterning. For a given SL potential, we define a pair of reciprocal lattice vectors, $\vec{G}_1$ and $ \vec{G}_2 $, then we perform a BZ folding to study the lattice system. We write the momentum $\vec{q}_{mn}$ within the graphene Brillouin zone into a momentum $\vec{k}$ within the boundaries of the sBZ and a contribution from the SL. We then define the momentum $\vec{k}$ inside the sBZ such that:
\begin{equation}
\vec{q}_{mn} = \vec{k} + (m,n)\cdot (\vec{G}_1, \vec{G}_2) \equiv \vec{k} + \vec{G}_{mn},
\label{eq: Gmn}
\end{equation}
where $\vec{G}_{mn}=m\vec{G}_1+n\vec{G}_2$ with $m,n$ integers. Each $\vec{G}_{mn}$ vector in the reciprocal space has six or four nearest neighbors, depending on whether the geometry of the SL is triangular, kagome or square. For example, in the triangular and kagome SLs, the $\vec{G}_{mn}$ vectors with modulus  $|\vec{G}_1|$, $|\vec{G}_1+2\vec{G}_2|$ and $|2\vec{G}_2|$, generate the first, second and third harmonic functions, respectively. For a square SL, however, the $\vec{G}_{mn}$ vectors with modulus $|\vec{G}_1|$, $|\vec{G}_1+\vec{G}_2|$ and $|2\vec{G}_1|$ generate the first, second and third harmonic functions, respectively. By considering the SL potential as a perturbation, the low-energy electronic structure of graphene/$V_{SL}$ is given by a Hamiltonian of the form
\begin{equation}
H=H_{0}+V_\text{SL},\label{eq:MainHamiltonian}
\end{equation}
where the first term represents the monolayer low-energy Hamiltonian in the vicinity of a single valley. The second term $V_\text{SL}$ describes the effect of the patterned gate. The Hamiltonian in Eq.~\eqref{eq:MainHamiltonian} in a plane wave basis can be written as:
\begin{align}
H & = H_0(\vec{q}_{mn})\otimes \mathbb{I}_N + V_{SL}   \notag \\
& = H_0(\vec{k})\otimes \mathbb{I}_N + H_0(\vec{G}_{mn})\otimes \mathbb{I}_N + V_{SL}\notag \\
& = H_0(\vec{k})\otimes \mathbb{I}_N + H_{SL},
\label{eq: hamprincipal}
\end{align}
where $H_{SL}$ contains all superlattice effects, $H_0(\vec{k})$ is the graphene Hamiltonian in the sBZ and $\mathbb{I}_N$ is an identity matrix with dimensions given by the number of reciprocal lattice vectors used in the calculations. We note that the above equation can be straightforwardly extended to a multilayer graphene. 

\subsection{Superlattice Potential}

In an experimental setup, as shown in Fig.~\ref{fig:Device}, the electrostatic modulations arising from a patterned gate are transferred to the electrons in graphene, resulting in a modulated periodic potential of the form~\cite{Guinea2010Band,Song2015Topological}
\begin{equation}
    V_\text{SL}\left(\vec{r}\right)=
v_{0}+\sum_{j} v_{SL}(\vec{G}_j)e^{i\vec{G}_j\cdot\vec{r}},
\label{eq: superpotential}
\end{equation} 
where $v_{SL}(\vec{G}_j)$ are the Fourier components of the SL potential and $v_0$ constant shift. In general, the shape of the patterning can be a hole or a cylinder~\cite{krix2023patterned} placed at some distance away from the graphene layer. These kinds of potentials can be modeled by using muffin-tin-like functions~\cite{park2008new} which can simulate the sharp edges of the patterning if enough Fourier components are considered. However, the short-frequency components, responsible for the sharp edges, only contribute to a high energy regions and therefore, a good low energy approximation is to only consider a few harmonics. In fact, we found that the periodic potentials in Fig.~\ref{fig: Onsite} induced by the patterning in Fig.~\ref{fig:Device}, can be accurately described if we consider a potential of the form, 
\begin{equation}
    V_{SL}(\boldsymbol{r})=v_{0}+\sum_{j}v_{SL}(\vec{G}_j)\cos\left(\boldsymbol{G}_{j}\cdot\boldsymbol{r}\right).
    \label{eq: SLCosines}
\end{equation}
with $j=2m$ and $m\neq0$ is a positive integer.\ The full monolayer low energy  Hamiltonian has then the form 
\begin{equation}
H=v_F\boldsymbol{k}\cdot\boldsymbol{\sigma}+ V_{SL}(\boldsymbol{r})\sigma_{0},   
\label{eq: HamiltonianPauli}
\end{equation}
where $\boldsymbol{\sigma} = \{\sigma_x, \sigma_y \}$ are the Pauli matrices with identity matrix $\sigma_0$.\ Because the periodic potential $V_{SL}(\boldsymbol{r})$ is even, the operation $\mathcal{C}_2\mathcal{T}$, with $\mathcal{C}_2$ inversion symmetry and $\mathcal{T}$ time reversal, is preserved. Therefore, the Dirac cones in the sBZ resulting from the periodic potential in Eq.~(\ref{eq: SLCosines}) are gapless.\ We note that cosine functions have been successfully used to describe the even scalar contributions of hBN substrates in graphene monolayers~\cite{Wallbank2013Generic,SanJose2014Spontaneous,Jung2015Origin} and graphene bilayers with dielectric SLs~\cite{ghorashi2023multilayer,zeng2024gate}.

\begin{figure*}[t]
\includegraphics[width=\textwidth]{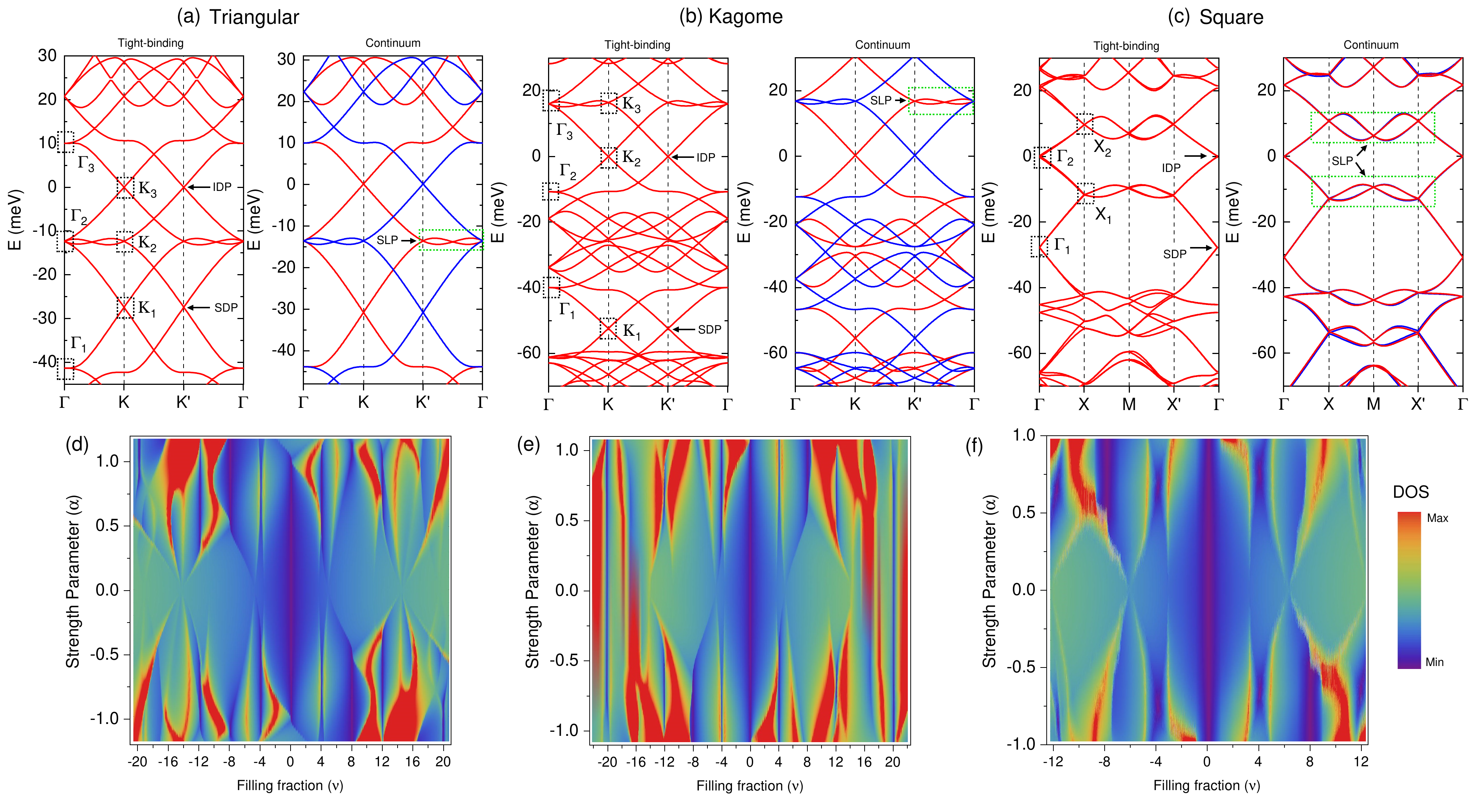}	
\caption{Band structure of graphene with (a) triangular, (b) kagome and (c) square SL potentials. In (d), (e) and (f) we show the corresponding DOS as a function of the strength parameter ($\alpha$) and the filling fraction ($\nu$) in the continuum model. Here, $\nu = \pm 4$ for a full/empty middle band. In the panels at the top row, the band structure on the left is calculated with a tight-binding model and on the right by an effective continuum model. All SL potentials have the same length $L_1=80$ nm, and SL gate bias $V_{si}=-80$ V. In the TB calculation, the scaling factor is $s_f=5$. In all panels the bands obtained by a continuum model are centered at the corresponding  $\vec{K}$ (red) and $\vec{K'}$ (blue) valleys. Green dashed square in each panel indicates the superlattice Dirac cones(SLP). The intrinsic Dirac point (IDP) and the satellite Dirac point (SDP) are illustrated by black arrows. The black rectangles outline the K points around the IDP, SLP and SDP.  
}
  \label{Fig: Figure4}
\end{figure*}

\subsection{Fourier Expansion of the SL Potential} \label{sec: FourierSL}

\begin{table}
\begin{tabular}{|c|c|c|c|}
\cline{2-4} \cline{3-4} \cline{4-4} 
\multicolumn{1}{c|}{} & $v_{1}$ (meV) & $v_{2}$ (meV) & $v_{3}$ (meV) \tabularnewline
\hline 
Triangular & 12.5 & 6.25 & 6.25\tabularnewline
\hline 
Kagome & 12 & -6 & -6\tabularnewline
\hline 
Square & 15 & 7.5 & 7.5\tabularnewline
\hline 
\end{tabular}
\caption{Fourier amplitudes for the different SL geometries using to simulated the device shown in Fig.~\ref{fig:Device}(a). 
}
\label{tb: table1}
\end{table}

In the TB model described in Sec.~\ref{sec: TB Model}, the effect of the patterned gate is introduced as a periodic on-site energy acting on the graphene sites. This method has the advantage that any periodic potential can be introduced independently of its geometry. In the continuum model, the Fourier expansion of the corresponding potential is required. To determine the number of Fourier harmonics, it is important to understand the experimental geometry of the patterning because the localization of the potential in real space determines the number of Fourier harmonics. In an experimental setup, shown in Fig.~\ref{fig:Device}, the holes are separated by some distance, and the flat regions between neighboring holes should also be considered. While a muffin-tin potential is a good option to mimic the sharp holes and the flat regions of a patterned gate~\cite{park2008new, barcons2022engineering}, it has the disadvantage of requiring special Bessel functions with a large number of harmonics. As mentioned before, we have found that a few harmonics with simple cosine functions provide a good description of the periodic potential. 

For both triangular and kagome lattices, the Fourier coefficients in Eq.~(\ref{eq: SLCosines}) runs over the first three harmonic functions. In the triangular SL, the first harmonic determines the shape of triangular potential, Fig.~\ref{fig: Onsite}(a) (top) while the second and third harmonics are require to flatten the regions between maxima [the region around $x=40$ nm in Fig.~\ref{fig: Onsite}(a)].\ In the kagome lattice, shown in the top panel of Fig.~\ref{fig: Onsite}(c), the first harmonic is required to simulate the triangular lattice within the hexagons, the second and third harmonics are required to simulate the shared corner triangles~\cite{Wang2024Dispersion}.\ Similarly, the square SL potential also requires three harmonics.\ The first one determines the shape of the potential, while the second and third harmonic is required to flatten the regions between maxima (see the bottom panel of Fig.~\ref{fig: Onsite}(b)). We note that this representation for a square lattice was used to describe the experiment in Ref.~\cite{forsythe2018band}.\ In Table~\ref{tb: table1}, we show the magnitudes of Fourier components obtained by numerically fitting the potential resulting from the simulated device in Fig.~\ref{fig: Onsite}(a).\ It is worth noting that while the TB model modifies the SL strength through the gate bias $V_{si}$ and the position of the Fermi level through $V_{g}$, the continuum model adjusts it by introducing a single modulation parameter $\alpha$, resulting in Fourier components given by $\alpha \{v_1,v_2,v_3\}$ where each coefficient corresponds to the amplitude of the first, second and third harmonic functions in Eq.~(\ref{eq: SLCosines}), respectively.

\section{Electronic structures}
\label{results_sec}

\subsection{Band Structure}
In Fig.~\ref{Fig: Figure4}, we display the band structure for three different SL geometries calculated by both the TB model and a low-energy continuum model. The bands obtained from the low-energy calculations are evaluated around the corresponding valley and then plotted in the common BZ~\cite{Moon2014Electronic} on the $k$-space path shown in Fig.~\ref{fig:FigLattice}(c). In this panel, we set the SL gate bias as $V_{si} = -80$ V, corresponding to $\alpha = \{0.85, 0.60, 1.0\}$ for triangular, kagome, and square lattices, respectively. In the TB calculation, the primitive lattice vectors of the triangular and kagome SL are $\vec L_1=65 \vec a_1$ and $\vec L_2=65 \vec a_2$, and of the square SL are $\vec L_1=63 \vec a_1$ and $\vec L_2'=37 \vec a_2'$. Figure~\ref{Fig: Figure4} demonstrates excellent agreement between the band structures calculated using the TB and continuum models, indicating that the Fourier expansion given by Eq.~(\ref{eq: SLCosines}) is an effective method to simulate the periodic potential induced by patterned devices. 

In the spectrum, while all three systems shown in Fig.~\ref{Fig: Figure4} possess the graphene intrinsic Dirac point (IDP), it is noticeable that the periodic potential induces different properties in the high energy bands depending on the geometry. We also illustrate the variation of the DOS as a function of the filling fraction ($\nu$) where $\nu=\pm4$ for fully/empty middle bands. An interesting feature, resulting from the periodic potential in the folded BZ, is the presence of superlattice Dirac cones (SLP), shown within green dashed squares in Fig.~\ref{Fig: Figure4}. In the triangular and kagome SL, these cones closely resemble those from a moiré modification on graphene due to a hBN substrate~\cite{ponomarenko2013cloning}. This feature has been experimentally observed in square~\cite{huber2020gate,forsythe2018band}, triangular~\cite{forsythe2018band}, and recently in monolayer graphene placed on a kagome patterned gate~\cite{Wang2024Dispersion}.\ In particular, by comparing the electronic structures of the triangular in Fig.~\ref{Fig: Figure4}(a) and kagome in Fig.~\ref{Fig: Figure4}(b), the SLP of the former are found in the negative energy region, while those of the latter are in the positive one. A closer inspection of the electronic structures and density plots reveals that in the low energy region and small $\alpha$, both kagome and triangular have a similar electronic structure if we invert the sign of $\alpha$ in one of them. The reason for this lies in the shape of the periodic potential for both geometries. As shown in Table~\ref{tb: table1}, the sign of the first harmonic functions, which control the triangular part of both potentials, have opposite signs. However, while a simple triangular SL can be described only with a first harmonic expansion~\cite{Guinea2010Band} [as described in Sec.~\ref{sec: FourierSL}, we added the second and third harmonics to mimic the flat regions in Fig.~\ref{fig: Onsite}(a)], a kagome potential is an intrinsically high-order potential requiring at least three periodicities to shape it~\cite{Wang2024Dispersion}. 

The electronic structure and the corresponding variation of the DOS  for a square lattice are shown in Fig.~\ref{Fig: Figure4}(c) and Fig.~\ref{Fig: Figure4}(f).\ The bands are similar to those obtained in Ref.~\cite{huber2020gate, forsythe2018band}, where the resistance maps revealed SLP above and below the main IDP. Our results in Fig.~\ref{Fig: Figure4}(f) are consistent with the experiments in Ref.~\cite{huber2020gate, forsythe2018band}, because the DOS minimum typically aligns with the resistance maximum in transport measurements. An additional Dirac cone, labeled as $\Gamma_1$ is obtained in a lower energy. In the high-energy regions, we found a discrepancy between the TB and continuum models. A thorough review of our numerical calculations revealed that, while the continuum model does not capture the lattice sites and implicitly assumes a perfect square geometry, the TB calculations cannot define a perfect square (see Sec.~\ref{subsec:Lattice}). The SL potential in Fig.~\ref{fig: Onsite}(b) acting on each graphene site is not a perfect square. This is the main reason for the discrepancies in the high-energy regions. However, in the low energy regions near IDP, SLP, and SDP in Fig.~\ref{Fig: Figure4}(c), the band structures are identical.

\begin{figure}[t]
\includegraphics[width=0.45\textwidth]{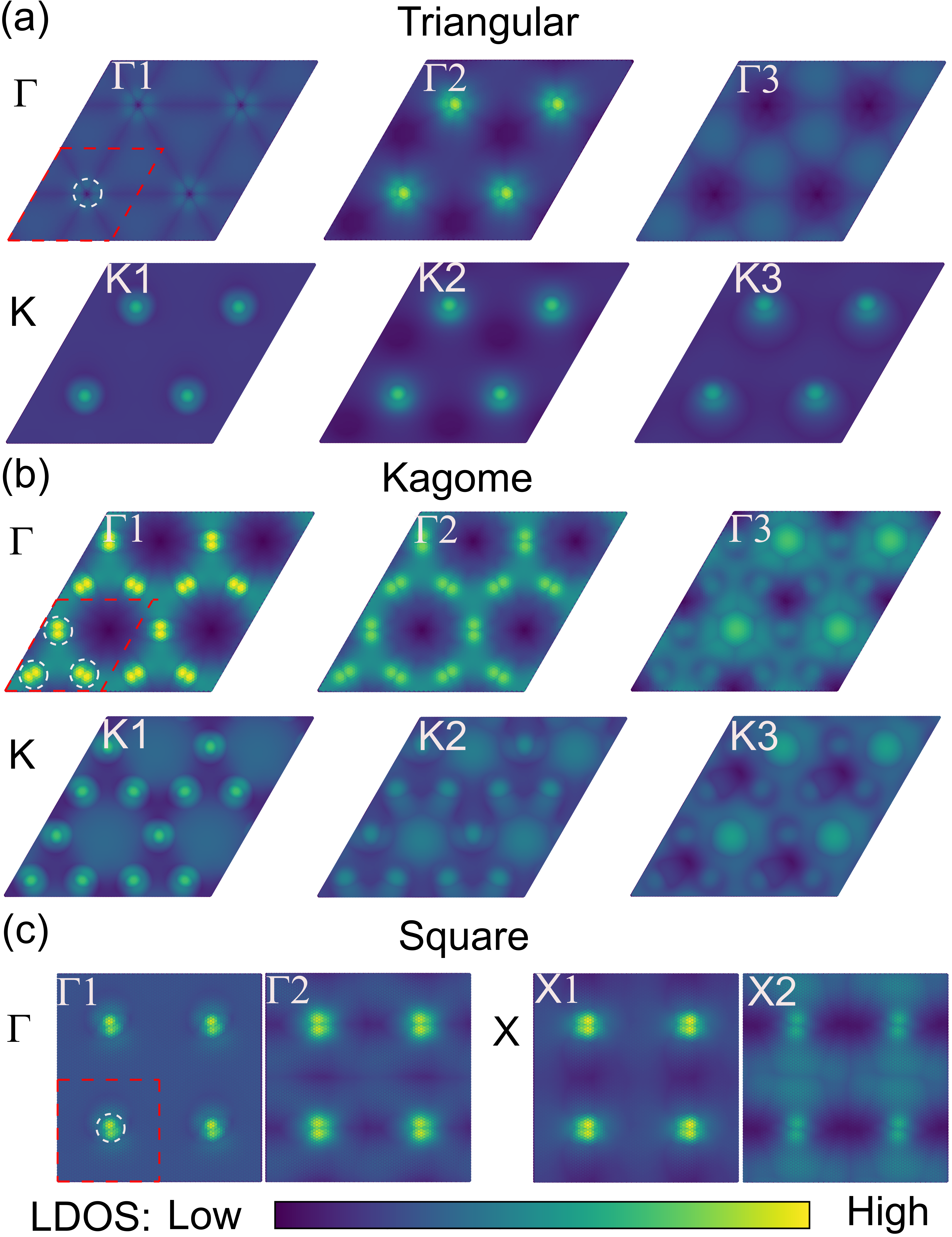}	
\caption{Distributions of the electronic states in real space corresponding to the different points in the sBZ. The K points of the sBZ are illustrated in Fig. \ref{Fig: Figure4}.
(a) For the triangular SL case, (b) for the kagome SL case, and (c) for the square case. The states are obtained by directly diagonalizing the TB Hamiltonian. The three SL potentials have the same length $L_1=80$ nm, and SL gate bias $V_{si}=-80$ V. The unit cell of the SL and the artificial-lattice are outlined by the red dashed lines and white dashed circles, respectively.}
  \label{Fig: ChargeDens}
\end{figure}

\begin{figure*}[t]
\includegraphics[width=\textwidth]{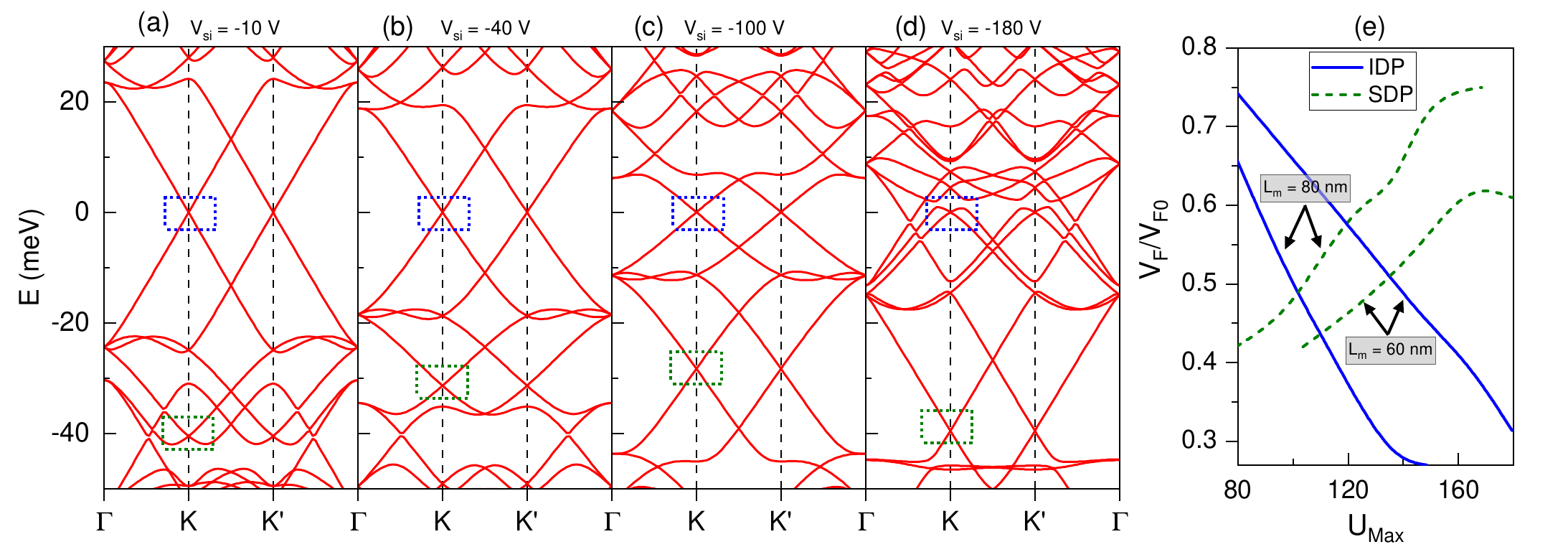}	
\caption{Electronic Band structure of graphene under a triangular SL potential with different SL gate biases $V_{si}$. Parameters are shown at the top of each panel. The corresponding potential strengths $U_{max}$ are $0.04$, $0.079$, $0.125$, and $0.168$ eV, respectively. The SL period is $L_1 = 80$ nm. The zero-energy points are fixed at the IDP of monolayer graphene. The band structures are calculated using a tight-binding model with a scaling factor $s_f = 5$. In panel (e), we compare the corresponding Fermi velocity, $v_F = \frac{1}{\hbar}\frac{\Delta E}{\Delta K}$, around the IDP (blue continuous line) and SDP (dashed green line) as a function of the SL potential strength for two values of the SL period, as indicated in the figure.}
  \label{Fig: Figure6}
\end{figure*}

We now consider the localization of the electronic states in real space. In Fig.~\ref{Fig: ChargeDens}(a) we show the electronic charge density of a triangular SL potential. The states of the K1, K2, $\Gamma$2 and K3 points, c.f. Fig.~\ref{Fig: Figure4}(a), are mainly localized in the patterned gate sites. The state at  $\Gamma$1 is nearly uniform in real space. The states of $\Gamma$3 do not have components in patterned gate sites, instead, they are distributed around them. In consequence, due to the distinct charge distribution, states from these high-symmetry points will have a different response to the SL potential. This may allow to engineer the electronic structures around these points. The case of the kagome potential is more complicated. As shown in Fig.~\ref{Fig: ChargeDens}(b), the states at K1, K2 and K3 have contributions from both kagome (patterned gate sites) and triangular regions (between patterned gate sites). In contrast, the states at $\Gamma$1 and $\Gamma$2 are only distributed at the kagome sites. The distribution of $\Gamma$3 and K3 have strong contributions from the triangular regions. These points are part of the SLP which results from the triangular Fourier component of the kagome SL potential. In the square SL, the charge density for the symmetry points shown in Fig.~\ref{Fig: ChargeDens}(c), are mainly localized in the patterned gate sites. In particular, the states at $\Gamma$1 have a uniform contribution from other regions. The resemblance between states in the different sBZ points for the square lattice will result in a similar response to the external potential, in consequence, the bands will be rigidly shifted with the potential strength.  

\subsection{SL Potential Strength}

The SL periodic potential is one tuning knob to engineering the band structures of graphene. We define the SL strength, $U_{\text{Max}}$, as the extreme value of the potential in the patterned gate sites. As shown in Fig.~\ref{fig: Onsite}, where the SL strength is the same in the three geometries and is controlled by the bottom gate voltage  $V_{si}$, which provide a way for experimental realization of \textit{in situ} gate-tunable band engineering. Figure~\ref{Fig: Figure6} shows the evolution of the band structure with the bottom gate voltage for a triangular superlattice. For a weak SL potential, $V_{si}=-10$ V, the features of the SDPs (green dashed square) are obscured by other states existing around the same energy range. As mentioned in the previous subsection, the different band components respond differently to the SL potentials. Regions with a charge density similar to the superlattice potential will be more sensitive than those with a different charge distribution. This is why, as the SL potential increases, the states at $\Gamma$2 and K2 are strongly shifted in energy. Consequently, as the SL potential strength increases, an energy window gradually opens, isolating the SDP, as shown for $V_{si}=-40$ V in Fig.~\ref{Fig: Figure6}(b). Moreover, as shown in Fig.~\ref{Fig: Figure6}(e), the Fermi velocity of the SDP gradually increases, whereas the Fermi velocity of the IDP gradually decreases. For a triangular SL, under the lowest order perturbation, a renormalization of the Fermi velocity close to the IDP is proportional to $U_{\text{Max}}^2$\cite{Guinea2010Band}. Therefore, the Fermi velocity of the graphene can be \textit{in situ} tuned via the bottom gate bias. This is a remarkable observation, since the Klein tunneling of the monolayer graphene depends on the Fermi velocity~\cite{katsnelson2006chiral}. We could realize a graphene system with dynamically tunable Klein tunneling by simply varying the SL gate bias, which is fundamentally important for transport properties. With the SL potential large enough, as in the case of $V_{si}=-180$ V in Fig.~\ref{Fig: Figure6}(d), the IDP is strongly distorted and drowned by other states. The same analysis can be performed for other geometries, but in general, we argue that if a particular state has a charge density similar to the superlattice potential, it will be very sensitive to changes in $V_{si}$. If a state has a charge density completely different from the SL potential, then it won't be modified. This simple rule can be used to determine if a particular band will be distorted by modifying the SL potential strength. 

\begin{figure}[t]
\includegraphics[scale = 0.35]{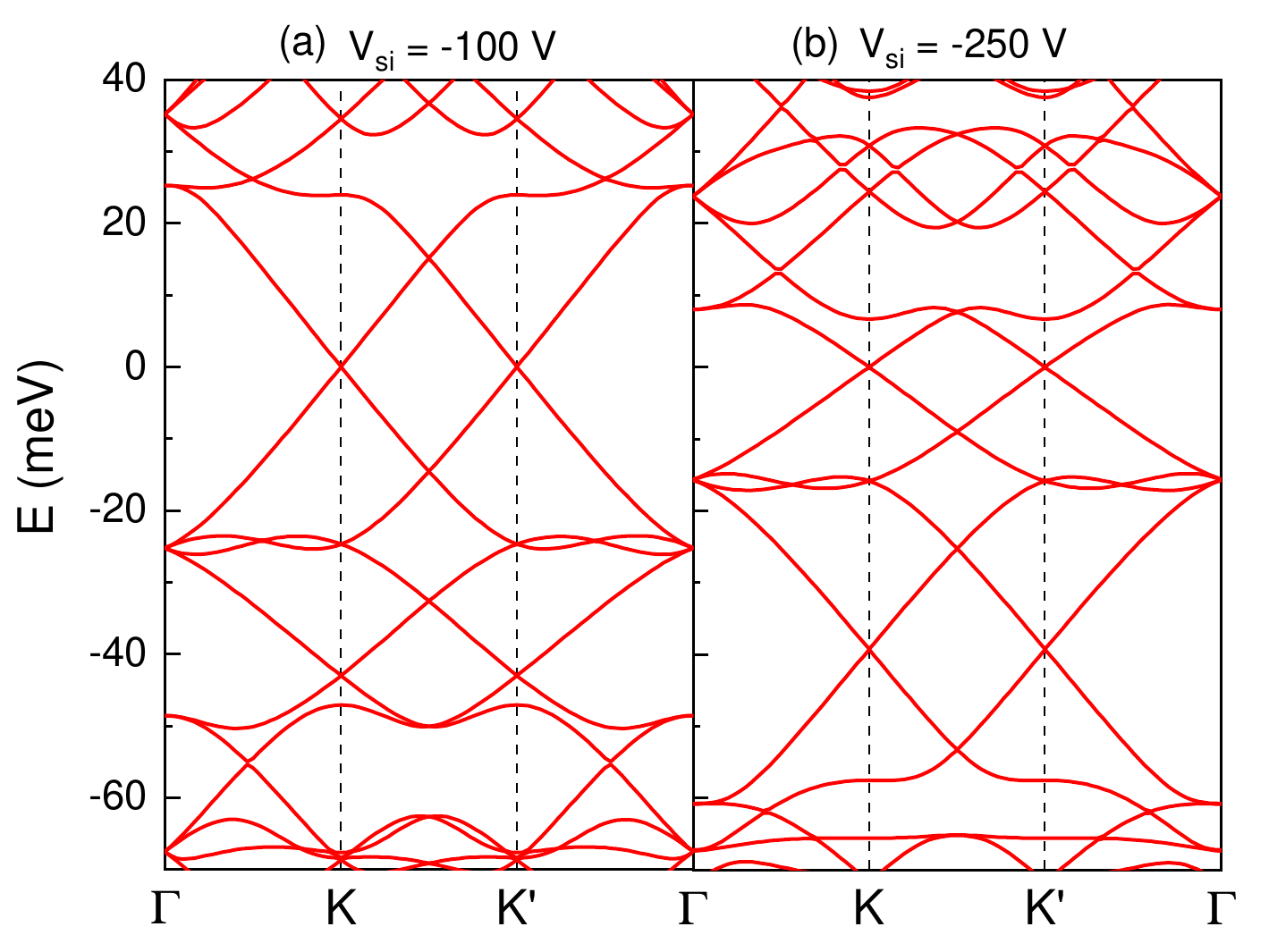}	
\caption{Band structures of graphene under a triangular SL potential with length $L_1=60$ nm and SL bias, (a) $V_{SL}=$ -100 V and (b) -250 V. The corresponding $U_{\text{Max}}$ are $0.103$ and $0.164$ eV, respectively. 
}\label{Fig: Figure7}
\end{figure}

\subsection{Relation between SL Period and SL Strength}

The SL period $L_m$ is another tuning knob to engineer band structures of graphene with patterned gates. The scaling properties of the Dirac equation of graphene imply that, if the dimension of the SL is reduced, $L_m$ $\to$ $\lambda L_m$, with $\lambda <1$, then an enhanced SL potential $U_{\text{Max}}$ $\to$ $\frac{1}{\lambda}U_{\text{Max}}$ leads to the same band structure with enlarged energies E $\to$ $\frac{1}{\lambda}$E. This scaling effect is clear if we compare Fig.~\ref{Fig: Figure6}(b) and Fig.~\ref{Fig: Figure7}(a), here, when the SL period is reduced from $L_m=80$ nm to $L_m=60$ nm, with $\lambda=0.75$, the potential strength is enhanced from $U_{\text{Max}}=0.079$ eV ($V_{si}=-40$ V) to $U_{\text{Max}}=0.103$ eV ($V_{si}=-100$ V). Similarly, we can compare Fig.~\ref{Fig: Figure6}(c) and Fig.~\ref{Fig: Figure7}(b), where the SL strength variation is from $U_{\text{Max}}=0.125$ eV ($V_{si}=-100$ V) to $U_{\text{Max}}=0.164$ eV ($V_{si}=-250$ V). In both cases, the band structures are similar but with different energy windows. Notably, for a SL potential with a larger period, the bottom gate $V_{si}$ introduces a stronger SL potential at the same gate bias. This is an important effect to consider when designing patterned gates. If the goal is to analyze effects due to remote bands, a larger SL period is beneficial. This technique has been used to access remote bands in monolayer graphene under a kagome SL~\cite{Wang2024Dispersion}.

\subsection{Electron-electron Interactions: Emergent Periodic Potentials}
In the previous section, we use a quantum capacitance model~\cite{liu2013theory,Fang2007Carrier} to simulate the potential introduced by a patterned gate.\ The solution of this model is equivalent to the self-consistent Poisson-Dirac method~\cite{liu2013theory}.\textbf{} The effect of the patterning introduce an effective scalar potential which acts as a Hartree-like periodic potential on monolayer graphene.\ To perform a fully self-consistent mean field calculation we now consider the effect of a Fock or exchange interaction. We follow the procedure described in Refs.~\citep{Cea2019Electronic,Cea2020Band,Pantaleon2021Narrow} by only considering those solutions where there is no spin or valley polarization. The total Hamiltonian is 

\begin{equation}
H=v_F\boldsymbol{k}\cdot\boldsymbol{\sigma}+ V_{SL}+V_{F},   
\label{eq: HamiltonianTotal}
\end{equation}
where the matrix elements of the exchange potential $V_\text{F}$ are given by
\begin{widetext}
\begin{equation}
v_{F,\boldsymbol{k}}\left(\boldsymbol G^{\prime}, \boldsymbol G\right)=-\sum_{\boldsymbol{q},l,i,j,\boldsymbol {G''}}\frac{V_{C}\left(\boldsymbol{q}-\boldsymbol{k}-\boldsymbol{G''}\right)}{\Omega}\phi_{\boldsymbol{q},l}^i\left(\boldsymbol{G^{\prime}}+\boldsymbol{G''}\right)\phi_{\boldsymbol{q},l}^{*,j}\left(\boldsymbol{G}+\boldsymbol{G''}\right),
\label{eq: vfock}
\end{equation}
\end{widetext}
with $V_{C}(q)$ the Fourier transform of the Coulomb potential. In Eq.~\eqref{eq: vfock}, $\Omega$ is the area of the real space unit cell, $i,j$ are sublattice indexes and $l$ runs over all occupied states above a given threshold. Here we set this threshold by considering up to six bands below charge neutrality point (for details please refer to Ref.~\citep{Cea2020Band}) and we only consider the effect at charge neutrality, where the Fock term is largest. The matrix elements in Eq.~\eqref{eq: vfock} for $\boldsymbol G = \boldsymbol G'$ are real numbers and they contribute as an on-site momentum dependent term. In twisted bilayer graphene, this term breaks inversion symmetry~\cite{Cea2020Band}. The non-diagonal terms with $\boldsymbol G \neq \boldsymbol G'$ are the Umklapp processes and they involve overlaps between the different components of the wavefunctions~\cite{Pantaleon2021Narrow}. 

\begin{figure*}
\includegraphics[scale = 0.43]{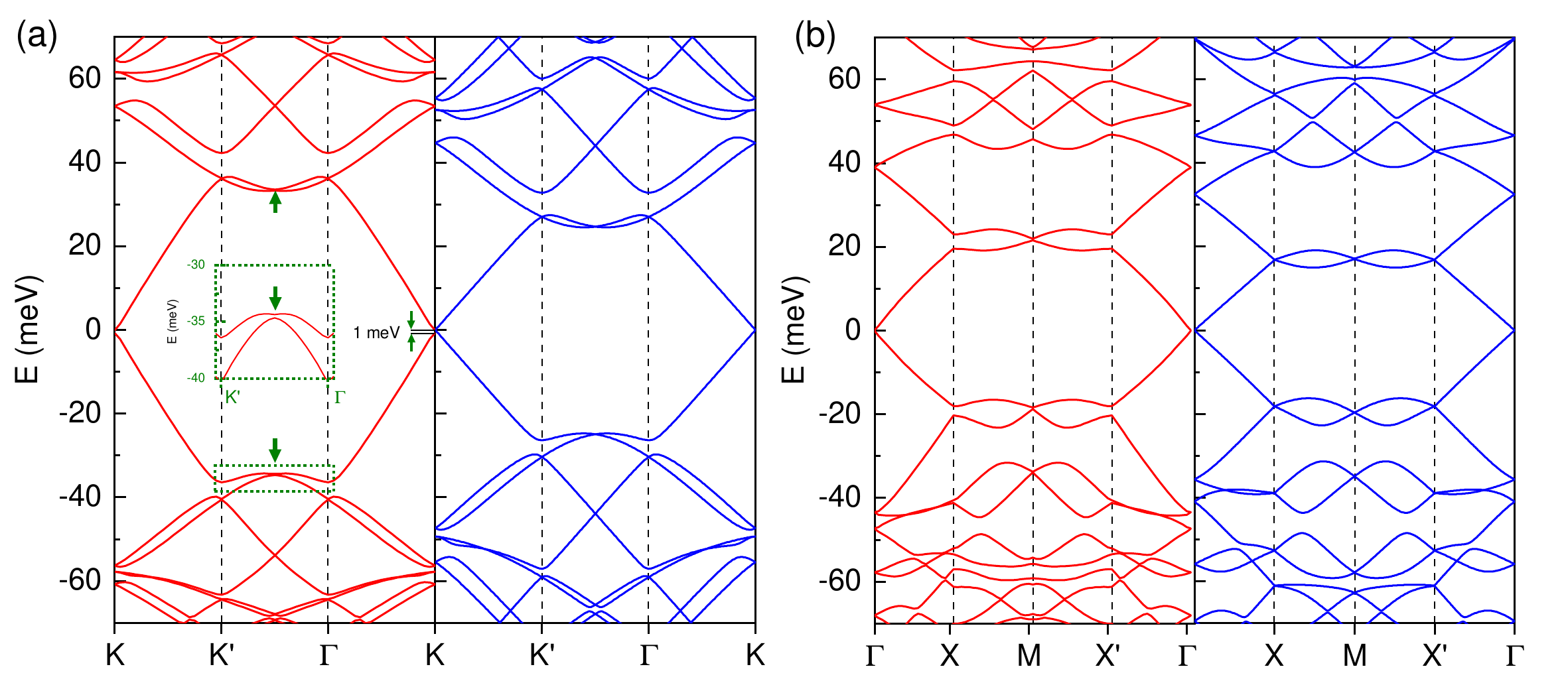}	
\caption{Interacting (red) and non-interacting (blue) band structure of graphene under (a) triangular ($\alpha = -0.25$) and (b) square ($\alpha = -0.016$) SL potential. In both cases the SL length is set to $L_1=80$ nm. Green arrows in (a) indicate the small gaps induced by the self-consistent Fock interaction. Green inset in (a) is an enlarged region where a gap with the remote band is indicated. }
\label{Fig: Figure8}
\end{figure*}

A comparison between the Fock and non-interacting bands is shown in Fig.~\ref{Fig: Figure8} and some interesting features are in order here.\ For a triangular SL in Fig.~\ref{Fig: Figure8}(a), as shown in the left panel, there is a renormalization of the Fermi velocity which is manifested by the change in the slope of the energy bands near $\Gamma$.\ This effect, previously reported in graphene monolayers~\cite{Kotov2012Electron,Stauber2017Interacting,Elias2011Dirac} implies that the Fock interaction is non-zero for the overlaps between the $i\neq j$ terms in Eq.~(\ref{eq: vfock}).\ Because the Fock potential is periodic in the SL unit cell, this suggest an emergent electronic gauge-like periodic potential that depends on the strength of the Coulomb interactions. We note that, if the dielectric constant (set to $\epsilon=4$) increases, all Fock effects are suppressed. In addition, we also found an emergent periodic mass potential resulting from the momentum-dependent diagonal terms with $\boldsymbol G = \boldsymbol G'$ in Eq.~(\ref{eq: vfock}).\ This emergent potential, with a strength between 2-4 meV, is strong enough to break inversion symmetry and thus opening a mass gap, as shown in Fig.~\ref{Fig: Figure8}.\ We verified the persistence of these effects with different values of $L_m$ and also for a kagome geometry.\ In the case of a square lattice, we also found an emergent periodic potential strong enough to open a gap at the $X$ and $X'$ points.\ We note that no mass gap is found at $\Gamma$ or $M$ which indicates that this potential does not break inversion symmetry as in the triangular case.\ In fact, by a numerical inspection of the Fock matrices we found that this potential is also gauge-like and, as in the triangular case, is originated by the $i \neq j$ terms in Eq.~(\ref{eq: vfock}).\ In addition, it is important to mention that for some seeds in the self-consistent calculations for the Fock interaction in the triangular superlattice, we found that the combined effect of electron-electron interaction and patterned gate potentials may give rise to an isolated Chern band with $\mathcal{C} = 1$.\ However, while such solutions are extremely sensitive to the parameters and initial conditions, we do not exclude their existence in realistic systems. Because a full phase diagram calculation, as a function of Coulomb strength, SL period, and SL strength, is required to determine their stability, we leave their analysis for future work. The presence of topological bands due to patterned gates has already been explored in graphene bilayers~\cite{ghorashi2023topological}.\ Our findings suggest that increasing the Coulomb strength by reducing the dielectric constant, modifying the superlattice period, gate distance, or adding additional layers may significantly impact electron-electron interactions and even induce additional topological phases~\cite{ghorashi2023topological, sun2023signature, ghorashi2023multilayer}.

\section{Conclusion} \label{conclu}
  
Using a quantum capacitance model approach, we simulate realistic devices capable of generating periodic potentials acting on graphene monolayers. We analyze the electronic bands of the resulting superlattice structure by considering three generic geometries: triangular, kagome, and square. We construct both TB and continuum models for these systems and find that only a few harmonics are required to fully describe the effect of the electrostatic gate on the graphene monolayer. Our procedure can be used to describe and simulate realistic devices with almost any periodic patterned gate geometry acting on graphene monolayers.  

Our findings reveal that the electronic spectra and charge density are strongly tunable by the SL strength. Notably, the different charge distributions at various points within the sBZ play a key role in the charge response to the electrostatic gate, making some regions of the bands more sensitive to the SL strength. We identify a simple relationship between the SL strength and the SL period, demonstrating how remote bands can be accessed with smaller voltages.

By introducing a self-consistent Fock interaction, we found an emergent periodic potential strong enough to break inversion symmetry and open a mass gap. This effect, highly dependent on electron-electron interactions, which combined with the SL potential, gives rise to a band with a non-zero Chern number. However, this effect is extremely sensitive to the parameters, and we believe it may be significant in graphene multilayers where the superlattice bands are narrower and isolated from the remote bands~\cite{ghorashi2023multilayer}.

\begin{acknowledgments}
We thank Francisco Guinea, Shengjun Yuan, Xiaodong Fan, Changgan Zeng, Vo Tien Phong, Alejandro Jimeno-Pozo, Hector Sainz-Cruz, Saul Herrera and Gerardo Naumis for fruitful discussions.\ IMDEA Nanociencia acknowledges support from the ‘Severo Ochoa’ Programme for Centres of Excellence in R\&D (CEX2020-001039-S/AEI/10.13039/501100011033). P.A.P. and Z.Z acknowledge support from NOVMOMAT, project PID2022-142162NB-I00 funded by MICIU/AEI/10.13039/501100011033 and by FEDER, UE as well as financial support through the (MAD2D-CM)-MRR MATERIALES AVANZADOS-IMDEA-NC. Z.Z acknowledges support from the European Union's Horizon 2020 research and innovation programme under the Marie-Sklodowska Curie grant agreement No 101034431.\\
ZZ and YL contributed equally to this work.
\end{acknowledgments}

\bibliographystyle{apsrev4-2}

%


\end{document}